\documentclass[prd,nobibnotes,superscriptaddress,nofootinbib,twocolumn]{revtex4-1}

\usepackage{amsmath,amsfonts,bm}
\usepackage{graphicx,multirow}

\begin{document}
\title{Anomaly-free $U(1)$ gauge symmetries in neutrino seesaw flavor models}

\author{ Lu\'is~M.~Cebola}
\email{luismcebola@ist.utl.pt} \affiliation{Departamento de F\'{\i}sica and Centro
de F\'{\i}sica Te\'orica de Part\'{\i}culas (CFTP), Instituto Superior T\'ecnico (IST),
Universidade de Lisboa, Avenida Rovisco Pais, 1049-001 Lisboa, Portugal}
\author{ David~Emmanuel-Costa}
\email{david.costa@ist.utl.pt} \affiliation{Departamento de F\'{\i}sica and Centro
de F\'{\i}sica Te\'orica de Part\'{\i}culas (CFTP), Instituto Superior T\'ecnico (IST),
Universidade de Lisboa, Avenida Rovisco Pais, 1049-001 Lisboa, Portugal}
\author{Ricardo~Gonz\'{a}lez~Felipe}
\email{ricardo.felipe@ist.utl.pt} \affiliation{Instituto Superior de
Engenharia de Lisboa - ISEL, Rua Conselheiro Em\'{\i}dio Navarro 1, 1959-007
Lisboa, Portugal} \affiliation{Departamento de F\'{\i}sica and Centro de F\'{\i}sica
Te\'orica de Part\'{\i}culas (CFTP), Instituto Superior T\'ecnico (IST),
Universidade de Lisboa, Avenida Rovisco Pais, 1049-001 Lisboa, Portugal}

\begin{abstract}
Adding right-handed neutrino singlets and/or fermion triplets to the
particle content of the Standard Model allows for the implementation of the
seesaw mechanism to give mass to neutrinos and, simultaneously, for the
construction of anomaly-free gauge group extensions of the theory. We
consider Abelian extensions based on an extra $U(1)_X$ gauge symmetry,
where $X$ is an arbitrary linear combination of the baryon number $B$ and
the individual lepton numbers $L_{e,\mu,\tau}$. By requiring cancellation
of gauge anomalies, we perform a detailed analysis in order to identify the
charge assignments under the new gauge symmetry that lead to neutrino
phenomenology compatible with current experiments. In particular, we study
how the new symmetry can constrain the flavor structure of the Majorana
neutrino mass matrix, leading to two-zero textures with a minimal extra
fermion and scalar content. The possibility of distinguishing different
gauge symmetries and seesaw realizations at colliders is also briefly
discussed.
\end{abstract}
 \maketitle

\section{Introduction}

Neutrino oscillation experiments have firmly established the existence of
neutrino masses and lepton mixing, implying that new physics beyond the
Standard Model (SM) is required to account for these observations (for
reviews, see e.g. Refs.~\cite{Strumia:2006db,Branco:2011zb}). One of the most
appealing theoretical frameworks to understand the smallness of neutrino
masses is the seesaw mechanism. In this context, the tree-level exchanges of
new heavy states generate an effective dimension-five Weinberg operator,
which then leads to an effective neutrino mass matrix at low energies. A
simple possibility consists of the addition of singlet right-handed neutrinos
(type I seesaw). Alternatively, color-singlet $SU(2)$-triplet scalars (type
II) or $SU(2)$-triplet fermions (type III) can be introduced.

As is well known, theories that contain fermions with chiral couplings to the
gauge fields suffer from anomalies, i.e. the breaking of gauge symmetries of
the classical theory at one-loop level. To make such theories consistent,
anomalies should be exactly canceled. In the SM, this cancellation occurs
between quarks and leptons within each
generation~\cite{Gross:1972pv,Bouchiat:1972iq,Georgi:1972bb}. Adding new
chiral fermions requires us to arrange the chiral sectors of the new theory
so as to cancel the arising gauge anomalies.

One attractive possibility is to realize the anomaly cancellation through the
modification of the gauge symmetry. Extra $U(1)$ symmetries naturally arise
in a wide variety of grand unified and string theories. One of the
interesting features of such theories is their richer phenomenology, when
compared with the SM (for reviews, see e.g.
Refs.~\cite{Leike:1998wr,Langacker:2008yv}). In particular, the spontaneous
breaking of additional gauge symmetries leads to new massive neutral gauge
bosons which, if kinematically accessible, could be detectable at the Large
Hadron Collider (LHC). Clearly, the experimental signatures of these theories
crucially depend on whether or not the SM particles have nontrivial $U(1)_X$
charges. Assuming that the SM fermions are charged under the new gauge group,
and that the new gauge boson $Z'$ has a mass around the TeV scale, one
expects some effects on the LHC phenomenology.

In the context of neutrino seesaw models, the implications of anomaly-free
constraints based on the gauge structure $SU(3)_{C}\otimes SU(2)_{L} \otimes
U(1)_{Y} \otimes U(1)_X$ have been widely studied in the
literature~\cite{Barr:1986hj,Ma:2001kg,Barr:2005je,Montero:2007cd,Adhikari:2008uc,EmmanuelCosta:2009za}.
In particular, assuming family universal charges, it was shown in
Ref.~\cite{EmmanuelCosta:2009za} that type I and type III seesaw mechanisms
cannot be simultaneously realized, unless the $U(1)_X$ symmetry is a replica
of the standard hypercharge or new fermionic fields are added to the theory.
On the other hand, when combined type I/II or type III/II seesaw models are
considered, it is always possible to assign nontrivial anomaly-free charges
to the fields. Models based on gauge symmetries that are linear combinations
of the baryon number $B$ and the individual lepton flavor numbers $L_\alpha$
($\alpha=e,\mu,\tau$) have also been extensively
discussed~\cite{Ma:1997nq,Ma:1998dp,Ma:1998zg,Chang:2000xy,Salvioni:2009jp,
Heeck:2012cd,Araki:2012ip}. From the phenomenological viewpoint many aspects
of the latter symmetries are similar to those of the $B-L$ symmetry, with
$L=\sum_\alpha L_\alpha$ being the lepton number. Nevertheless, the flavor
information encoded in the new gauge symmetry can lead to definite
predictions on the neutrino mass spectrum and mixing.

In this paper, we consider Abelian extensions of the SM based on an extra
$U(1)_X$ gauge symmetry, with $X \equiv a\, B - \sum_{\alpha} b_\alpha
L_\alpha$ being an arbitrary linear combination of the baryon number $B$ and
the individual lepton numbers $L_\alpha$. Our aim is to perform a systematic
study, thus complementing previous works in several aspects. In particular,
we classify all the anomaly-free $U(1)_X$ gauge symmetries that lead to
predictive two-zero textures in the effective neutrino mass matrix, obtained
via type I, type III or mixed type I/III seesaw mechanisms, with a minimal
extra matter content. Some of these symmetries have been recently identified
in Ref.~\cite{Araki:2012ip} in a type I seesaw framework. We extend the
analysis and obtain new solutions not previously found in the literature.
Furthermore, we study in detail the phenomenological implications of such
symmetries on neutrino oscillation analysis through matter effects. Finally,
we also discuss the possibility of discriminating at collider experiments the
flavor structure of the neutrino mass matrix and its corresponding gauge
symmetry, by studying the decays of the $Z'$ boson into leptons and
third-generation quarks.

The paper is organized as follows. In Sec.~\ref{sec:anomalyfree}, by
requiring cancellation of gauge anomalies, we study the allowed charge
assignments under the new gauge symmetry, when two or three right-handed
neutrino singlets or fermion triplets are added to the SM particle content.
In Sec.~\ref{sec:pheno} we then discuss the phenomenological constraints on
these theories, requiring consistency with current neutrino oscillation data.
In particular, by extending the SM with a minimal extra fermion and scalar
content, we study how the new gauge symmetry can constrain the flavor
structure of the effective neutrino mass matrix, obtained through a type I or
type III seesaw mechanism. The possibility of distinguishing different charge
assignments (gauge symmetries) and seesaw realizations at collider
experiments is also briefly addressed. Our conclusions are summarized in
Sec.~\ref{sec:summary}.

\section{Anomaly constraints}
\label{sec:anomalyfree}

We consider a renormalizable theory containing the SM particles plus a
minimal extra fermionic and scalar content, so that light neutrinos acquire
seesaw masses. We include singlet right-handed neutrinos $\nu_R$ and
color-singlet $SU(2)$-triplet fermions $\Sigma$ to implement type-I and
type-III seesaw mechanisms, respectively. Besides the SM Higgs doublet $H$
that gives masses to quarks and leptons, a complex scalar singlet field $S$
is introduced in order to give Majorana masses to $\nu_R$ and $\Sigma$.

We assume that each fermion field $f$ has a charge $x_f$ under the new
$U(1)_X$ gauge symmetry. We work in a basis of left- and right-handed
fermions: $q_L \equiv (u_L, d_L)^T, \ell_L \equiv (\nu_L, e_L)^T, u_R, d_R,
e_R, \nu_R$. For quarks, a family universal charge assignment is assumed,
while leptons are allowed to have nonuniversal $X$ charges. To render the
theory free of the $U(1)_X$ anomalies, the following set of constraints must
then be satisfied:
\begin{widetext}
\begin{align} \label{eqanomaly}
\begin{split}
 A_1=&n_G\left( 2\,x_{q}-x_{u}-x_{d}\right) =0\,,\\
 A_2=&\frac{3n_G}{2}x_{q}+\frac12\sum^{n_G}_{i=1} x_{\ell i}
 \,-\,2\,\sum^{n_\Sigma}_{i=1} x_{\sigma i} =0\,,
\\
 A_3=&n_G\left(\frac{x_{q}}{6}\,-\,\frac{4x_{u}}{3} \,-\,\frac{x_{d}}{3}
\right)\, +\,\sum^{n_G}_{i=1}\left( \frac{x_{\ell i}}{2}\,-\, \,x_{e
i}\right)=0\,,
\\
A_4=&n_G\left(x^2_{q}\,-\,2x_{u}^{2} \,+\,x_{d}^{2}\right)
\,+\,\sum^{n_G}_{i=1}\left( -x_{\ell i}^{2}\,+\,x_{e i}^{2}\right) =0\,,
\\
A_5=&n_G\left(6\,x_{q}^3-3\,x_{u}^3-3\,x_{d}^3 \right)
+\sum^{n_G}_{i=1}\left(2\,x_{\ell i}^3-x_{e i}^3\right) -\sum_{i=1}^{n_R}x_{\nu
i}^3\,-\,3\sum_{i=1}^{n_\Sigma}x_{\sigma i}^3=0\,,
\\
A_6=&n_G\left(6x_{q}-3x_{u}-3x_{d} \right)\,+\,\sum^{n_G}_{i=1}\left(
2\,x_{\ell i}-x_{e i}\right)-\sum_{i=1}^{n_R}x_{\nu i}-3\sum_{i=1}^{n_\Sigma} x_{\sigma i}=0\,,
\end{split}
\end{align}
\end{widetext}
where $n_G=3$ is the number of generations, $n_R$ is the number of
right-handed neutrinos and $n_\Sigma$ is the number of $SU(2)$-triplet
fermions.\footnote{For a type I (type III) seesaw mechanism alone,
consistency with neutrino oscillation data requires $n_R \geq 2$ ($n_\Sigma
\geq 2$). Aside from this constraint, the number of right-handed neutrinos
(fermion triplets) is arbitrary. If both seesaws are simultaneously allowed,
cases with $n_R \geq 2$  and $n_\Sigma \geq 1$, or $n_R \geq 1$  and
$n_\Sigma \geq 2$, are viable as well.} The equations for $A_1, \ldots,A_5$
arise from the requirement of the cancellation of the axial-vector
anomalies~\cite{Adler:1969gk}, while the equation for $A_6$ results from the
cancellation of the mixed gravitational-gauge
anomaly~\cite{Delbourgo:1972xb}.

\begin{table*}[t]
\begin{tabular}{|c|c|c|c|}
  \hline
  $\,n_R\,$ & $\,n_\Sigma\,$ & \,Anomaly constraints\, & Symmetry generator $X$ \\ \hline
  2 & 0 & $b_i+b_j=3a,\,b_k=0$& $B-3L_j-b_i'(L_i-L_j)$\\
   &  & $b_i+b_j=0,\,b_k=0$& $L_i-L_j$\\
  0 & 2 & $b_i+b_j=0,\,b_k=0$ & $L_i-L_j$\\
  2 & 1 & $b_i+b_j=3a,\,b_k=0$& $B-3L_j-b_i'(L_i-L_j)$\\
   &  & $b_i+b_j=0,\,b_k=0$& $L_i-L_j$\\
  1 & 2 & $b_i+b_j=0,\,b_k=3a$ & $B-3L_k-b_i'(L_i-L_j)$\\
    &  & $b_i+b_j=0,\,b_k=0$& $L_i-L_j$\\
  3 & 0 & $b_i+b_j+b_k=3a$ & $\quad(B-L)+(1-b_i')(L_i-L_j)+(1-b_k')(L_k-L_j)\quad$ \\
   &  & $b_i+b_j+b_k=0$ & $-b_i'(L_i-L_k)-b_j'(L_j-L_k)$ \\
  0 & 3 & $b_i+b_j=0,\,b_k=0$& $L_i-L_j$ \\
  3 & 1 & $b_i+b_j=3a,\,b_k=0$& $B-3L_j-b_i'(L_i-L_j)$\\
   &  & $b_i+b_j=0,\,b_k=0$& $L_i-L_j$\\
  1 & 3 & $b_i+b_j=0,\,b_k=0$& $L_i-L_j$\\
  2 & 2 & $b_i+b_j=0,\,b_k=0$&$L_i-L_j$ \\
  \hline
\end{tabular}
\caption{\label{table1}Anomaly-free solutions for minimal type I and/or type
III seesaw realizations and their symmetry generators. In all cases, $i \neq j
\neq k$ and $b_i'\equiv b_i/a$. Cases with $a=0$ correspond to a purely
leptonic symmetry.}
\end{table*}

Since the anomaly equations are nonlinear and contain many free parameters,
some assumptions are usually made to obtain simple analytic solutions. For
instance, in family universal models, universal charges are assigned so that
anomaly cancellation is satisfied within each family. Family universality is
nevertheless not necessarily required and nonuniversal solutions can be
equally found~\cite{Liu:2011dh}. Assuming, for instance, a nonuniversal
purely leptonic gauge symmetry with $x_{\ell i}=x_{e i}$ and $n_R = n_\Sigma
= n_G =3$, the anomaly equations~\eqref{eqanomaly} lead to the following
charge constraints:
\begin{align}\label{eqexample}
\begin{split}
  &x_{e 1}+x_{e 2}+x_{e 3} = 0,\\
  &x_{\nu 1}+x_{\nu 2}+x_{\nu 3}=0, \\
  &x_{\sigma 1}+x_{\sigma 2}+x_{\sigma 3}=0,\\
    &x_{e 1}x_{e 2}x_{e 3}-x_{\nu 1}x_{\nu 2}x_{\nu 3}
    - 3\,x_{\sigma 1}x_{\sigma 2}x_{\sigma 3}=0.
  \end{split}
\end{align}
This system of equations has an infinite number of integer solutions. For
example, with the charge assignment $(x_{\ell 1}, x_{\ell 2}, x_{\ell
3})=(x_{e 1}, x_{e 2}, x_{e 3})=(1, 2, -3)$, one can have the solutions
$(x_{\nu 1},x_{\nu 2},x_{\nu 3})=(-1,-3,4)$ and $(x_{\sigma 1},x_{\sigma
2},x_{\sigma 3})=(1,2, -3)$, or $(x_{\nu 1},x_{\nu 2},x_{\nu 3})=(1,3,-4)$
and $(x_{\sigma 1},x_{\sigma 2},x_{\sigma 3})=(-1,-1, 2)$, among many others.

In this work, we shall consider models where $X \equiv a\, B -
\sum_{i=1}^{n_G} b_i L_i$ is an arbitrary linear combination of the baryon
number $B$ and individual lepton numbers $L_i$, simultaneously allowing for
the existence of right-handed neutrinos and fermion triplets that participate
in the seesaw mechanism to generate Majorana neutrino masses. Under the gauge
group $U(1)_{X}$, the charge for the quarks $q_L, u_R, d_R$, is universal,
$x_q=x_u=x_d=a/3$, while the charged leptons $\ell_{Li}, e_{Ri}$ have the
family nonuniversal charge assignment $x_{\ell i} = x_{e i}=-b_i$, with all
$b_i$ different. The latter condition guarantees that the charged lepton mass
matrix is always diagonal (i.e. it is defined in the charged lepton flavor
basis), assuming that the SM Higgs is neutral under the new gauge symmetry.
The right-handed neutrinos $\nu_{R}$ and/or the triplets $\Sigma$ are allowed
to have any charge assignment $-b_k$, where $k=1 \ldots n_G$.

Substituting the $U(1)_X$ charge values into the anomaly
equations~\eqref{eqanomaly}, we obtain the constraints
\begin{align}\label{eqcharges}
\begin{split}
  &\sum_{k \leq n_\Sigma} b_k =0,\\
  &\sum_{i=1}^{n_G} b_i = \sum_{j\leq n_R} b_j = n_G\, a,\\
  &\sum_{i=1}^{n_G} b_i^3-\sum_{j\leq n_R} b_j^3-3\sum_{k\leq n_\Sigma} b_k^3=0.
  \end{split}
\end{align}

The solutions of this system of equations and the corresponding symmetry
generators $X$ are presented in Table~\ref{table1}, for minimal type I and
type III seesaw realizations with $n_R +n_\Sigma \leq 4$. We note that in the
absence of right-handed neutrinos only purely leptonic ($a=0$) gauge symmetry
extensions are allowed. This is a direct consequence of the second constraint
in Eq.~\eqref{eqcharges}. Given the charge assignments, one can identify the
maximal gauge group corresponding to each solution. For instance, when
$n_R=3$ and $n_\Sigma=0$, the maximal anomaly-free Abelian gauge group
extension is $U(1)_{B-L}\times U(1)_{L_e-L_\mu}\times U(1)_{L_\mu-L_\tau}$,
as recently pointed out in Ref.~\cite{Araki:2012ip}.

In the next section, we shall study the phenomenological implications of
different anomaly-free $U(1)_X$ gauge symmetries on the flavor structure of
the effective neutrino mass matrix, assuming that the new symmetries lead to
texture zeros that are consistent with the present neutrino oscillation data
(see Appendix~\ref{appexA}). In particular, several neutrino matrix patterns,
with a maximum of two independent
zeros~\cite{Frampton:2002yf,Xing:2002ta,Xing:2002ap}, turn out to be
compatible with all data at $1\sigma$ level~\cite{Fritzsch:2011qv}.

\section{Phenomenological constraints}
\label{sec:pheno}

\subsection{Neutrino mass matrix and texture zeros from the gauge symmetry}
\label{sec:gaugesym}

Realistic gauge theories should include mechanisms to explain fermion masses.
This is achieved in the SM by introducing gauge-invariant Yukawa
interactions, which give rise to nonvanishing quark and charged lepton masses
once the Higgs field acquires a vacuum expectation value (VEV). Neutrinos, on
the other hand, are massless in the SM, so new physics is required to account
for their nonzero masses. An attractive and economical framework is the
seesaw mechanism.

\begin{table*}[t]
\begin{tabular}{|c|c|c|c|}
  \hline
   \ Symmetry generator $X$\ &$\quad |x_s|\quad$ & $\quad\mathbf{M}_{R}\quad$
   & $\quad\mathbf{m}_\nu\quad$\\
  \hline
   $B+L_e-L_\mu-3L_\tau$ & 2 &$\mathbf{D}_2$ & \multirow{4}{*}{$\mathbf{A}_1$}\\
   \cline{1-3}
  $B+3L_e-L_\mu-5L_\tau$ & 2 &
  \multirow{3}{*}{$\quad(\mathbf{M}_{R})_{11}=(\mathbf{M}_{R})_{23}=(\mathbf{M}_{R})_{33}=0\quad$}
   & \\
    $B+3L_e-6L_\tau$ & 3 &   & \\
    $B+9L_e-3L_\mu-9L_\tau$ & 6 &   & \\
   \hline
   $B+L_e-3L_\mu-L_\tau$ & 2 &$\mathbf{D}_1$ & \multirow{4}{*}{$\mathbf{A}_2$}\\
   \cline{1-3}
  $B+3L_e-5L_\mu-L_\tau$ & 2 &
  \multirow{3}{*}{$\quad(\mathbf{M}_{R})_{11}=(\mathbf{M}_{R})_{22}=(\mathbf{M}_{R})_{23}=0\quad$}
   & \\
    $B+3L_e-6L_\mu$ & 3 &   & \\
    $B+9L_e-9L_\mu-3L_\tau$ & 6 &   & \\
    \hline
 $B-L_e+L_\mu-3L_\tau$ & 2 &$\mathbf{B}_4$ & \multirow{4}{*}{$\mathbf{B}_3$}\\
   \cline{1-3}
  $B-L_e+3L_\mu-5L_\tau$ & 2 &
  \multirow{3}{*}{$\quad(\mathbf{M}_{R})_{13}=(\mathbf{M}_{R})_{22}=(\mathbf{M}_{R})_{33}=0\quad$}
   & \\
   $B+3L_\mu-6L_\tau$ & 3 &   & \\
    $B-3L_e+9L_\mu-9L_\tau$ & 6 &   & \\
  \hline
   $B-L_e-3L_\mu+L_\tau$ & 2 &$\mathbf{B}_3$ & \multirow{4}{*}{$\mathbf{B}_4$}\\
   \cline{1-3}
  $B-L_e-5L_\mu+3L_\tau$ & 2 &
  \multirow{3}{*}{$\quad(\mathbf{M}_{R})_{12}=(\mathbf{M}_{R})_{22}=(\mathbf{M}_{R})_{33}=0\quad$}
   & \\
    $B-6L_\mu+3L_\tau$ & 3 &   & \\
    $B-3L_e-9L_\mu+9L_\tau$ & 6 &   & \\
   \hline
\end{tabular}
\caption{\label{table2}Anomaly-free $U(1)$ gauge symmetries that lead to
phenomenologically viable two-zero textures of the neutrino mass matrix
$\mathbf{m}_\nu$ in a type I seesaw framework with 3 right-handed neutrinos.
In all cases, the Dirac-neutrino mass matrix $\mathbf{m}_D$ is diagonal and
the charge assignment $x_{\nu i} = x_{\ell i} = x_{e i}=-b_i$ is verified.
The solutions belong to the permutation set $\mathcal{P}_1$ (see
Appendix~\ref{appexA} for the classification). For a mixed type I/III seesaw
scenario with $n_R=3$ and $n_\Sigma=1$ only the solutions with $|x_s|=3$ remain
viable.}
\end{table*}

The Lagrangian terms that generate the SM fermion masses and are compatible
with (minimal) type I and type III seesaw models for Majorana neutrinos are
given by
\begin{equation}
\begin{split}
&\mathbf{Y}_{u}\,\overline{q_L} u_{R}\widetilde{H} +\mathbf{Y}_{d}\,\overline{q_L}
d_{R}H + \mathbf{Y}_{e}\,\overline{\ell_L} e_{R} H
+\mathbf{Y}_{\nu}\,\overline{\ell_{L}}\nu_{R} \widetilde{H}\\
&+\frac12 \mathbf{m}_R \nu_{R}^{T}C\nu_{R}+\mathbf{Y}_{1}\,\nu_{R}^{T}C\nu_{R} S +
\mathbf{Y}_{2}\,\nu_{R}^{T}C\nu_{R} S^\ast\\
&+ \frac12 \mathbf{m}_\Sigma \mathrm{Tr } \left(\Sigma^{T}C\,\Sigma\right)
+ \mathbf{Y}_{T}\,\overline{\ell_{L}}i\tau_{2}\Sigma\, H\,\\
&+\mathbf{Y}_{3}\,\mathrm{Tr } \left(\Sigma^{T}C\,\Sigma\right) S
+\mathbf{Y}_{4}\,\mathrm{Tr } \left(\Sigma^{T}C\,\Sigma\right) S^\ast
+\text{H.c.}, \label{Ly}
\end{split}
\end{equation}
where $H= (H^\dagger\, H^0)^T$ is the SM Higgs doublet, assumed neutral under
the $U(1)_X$ gauge symmetry, $\tilde{H} = i \sigma_2 H^\dagger$, and $S$ is a
complex singlet scalar field with a $U(1)_X$ charge equal to $x_s$. Here
$\mathbf{Y}_{u,d,e}$, $\mathbf{Y}_{\nu}$ and $\mathbf{Y}_{T}$ are $n_G\times
n_G$, $n_G \times n_R$, and $n_G \times n_\Sigma$ Yukawa complex matrices,
respectively; $m_R$ and $\mathbf{Y}_{1,2}$ are $n_R \times n_R$ symmetric
matrices, while $m_\Sigma$ and $\mathbf{Y}_{3,4}$ are $n_\Sigma \times
n_\Sigma$ symmetric matrices.

Notice that, in general, the $U(1)_X$ symmetry does not forbid bare Majorana
mass terms for the right-handed neutrinos and fermion triplets. For matrix
entries with $X=0$, such terms are allowed. In turn, entries with $X\neq0$
are permitted in the presence of the singlet scalar $S$, charged under
$U(1)_X$. The latter gives an additional contribution to the Majorana mass
terms once it acquires a VEV.

Since a universal $U(1)_X$ charge is assigned to quarks, the new gauge
symmetry does not impose any constraint on the quark mass matrices. However,
our choice of a nonuniversal charge assignment, with $x_{\ell i} = x_{e
i}=-b_i$ and all $b_i$ different, forces the charged lepton mass matrix to be
diagonal, so leptonic mixing depends exclusively on the way that neutrinos
mix. The effective neutrino mass matrix $\mathbf{m}_\nu$ is obtained after
integrating out the heavy right-handed neutrinos and fermion triplets. In the
presence of both (type I and type III) seesaw mechanisms it reads as
\begin{align}
\mathbf{m}_\nu \simeq -\mathbf{m}_D\ \mathbf{M}_R^{-1}\ \mathbf{m}_D^T
- \mathbf{m}_T\ \mathbf{M}_\Sigma^{-1}\ \mathbf{m}_T^T\,,
\end{align}
where
\begin{align}
\begin{split}
\mathbf{m}_D &= \mathbf{Y}_\nu \langle H \rangle, \quad
\mathbf{M}_R = \mathbf{m}_R + 2 \mathbf{Y}_{1} \langle S \rangle + 2
\mathbf{Y}_{2} \langle S^\ast \rangle,\\
\mathbf{m}_T &= \mathbf{Y}_T \langle H \rangle, \quad
\mathbf{M}_\Sigma = \mathbf{m}_\Sigma
+ 2 \mathbf{Y}_{3} \langle S \rangle + 2 \mathbf{Y}_{4} \langle S^\ast
\rangle.
\end{split}
\end{align}

In what follows we restrict our analysis to minimal seesaw scenarios with
$n_R +n_\Sigma \leq 4$. The requirement that charged leptons are diagonal
($b_1 \neq b_2 \neq b_3$) imposes strong constraints on the matrix textures
of $\mathbf{m}_D$ and $\mathbf{m}_T$. Indeed, considering either a type I or
a type III seesaw framework, only those matrices with a single nonzero
element per column are allowed. Furthermore, matrices with a null row or
column are excluded since they lead to a neutrino mass matrix with
determinant equal to zero, not belonging to any pattern with only two
independent zeros.\footnote{Mixed type I/III seesaw mechanisms can relax this
constraint.}

We look for anomaly-free $U(1)$ gauge symmetries that lead to
phenomenologically viable two-zero textures of the neutrino mass matrix
$\mathbf{m}_\nu$ (i.e. to patterns $\mathbf{A}_{1,2}$, $\mathbf{B}_{1,2,3,4}$
and $\mathbf{C}$ given in Appendix~\ref{appexA}). Solutions were found only
within a type I seesaw framework with three right-handed neutrinos, or in a
mixed type I/III seesaw scenario with three right-handed neutrinos and one
fermion triplet. Table~\ref{table2} shows the allowed solutions, for the
cases when the Dirac-neutrino mass matrix $\mathbf{m}_D$ is diagonal, which
implies the charge assignment $x_{\nu i} = -b_i$. All the solutions belong to
the permutation set $\mathcal{P}_1$ [see Eq.~\eqref{psets}]. We remark that,
for each pattern of $\mathbf{m}_\nu$, there are another 20 solutions
corresponding to matrices $\mathbf{m}_D$ with 6 zeros (i.e. permutations of
the diagonal matrix) and their respective charge assignments. Thus, all
together there exist 96 viable solutions. No other anomaly-free solutions are
obtained in our minimal setup.\footnote{Solutions leading to $\mathbf{M}_R =
\mathbf{D}_1, \mathbf{D}_2, \mathbf{B}_3, \mathbf{B}_4$ have been recently
considered in Ref.~\cite{Araki:2012ip}.} For a mixed type I/III seesaw with
$n_R=3$ and $n_\Sigma=1$, only the set of solutions with $|x_s|=3$ in
Table~\ref{table2} are allowed, since the anomaly equations imply that the
$b_k$ coefficient associated to the fermion triplet charge is always zero.

Notice also that, starting from any pattern given in Table~\ref{table2},
other patterns in the table can be obtained by permutations of the charged
leptons. For instance, starting from the symmetry generators that lead to the
$\mathbf{A}_1$ pattern, those corresponding to $\mathbf{A}_2$ and
$\mathbf{B}_3$ are obtained by $\mu \leftrightarrow \tau$ and $e
\leftrightarrow \mu$ exchange, respectively. Similarly, the $\mathbf{B}_4$
texture can be obtained from $\mathbf{A}_2$ through the $e \leftrightarrow
\tau$ exchange.

\subsection{Scalar masses and mixing}

We assume a minimal scalar content: one SM Higgs doublet $H$, neutral under
$U(1)_X$, and a complex singlet scalar $S$ with the charge assignment $x_s$
under $U(1)_X$. The VEV of the scalar $S$ breaks this symmetry spontaneously,
giving a contribution to the masses of the right-handed neutrinos and fermion
triplets. The scalar potential reads as
\begin{align}
\begin{split}
V=& -m^2\, H^\dagger H + \lambda (H^\dagger H)^2 - m_S^2\, S^\dagger S +
\lambda_S\, (S^\dagger S)^2\\
& + \beta\, (S^\dagger S)(H^\dagger H)\,,
\end{split}
\end{align}
with $m^2 >0$ and $m_S^2 >0$ to generate the VEVs $\langle H\rangle =
v/\sqrt{2}$\, ($v \simeq 246$~GeV) and $\langle S \rangle =v_S/\sqrt{2}\,$;
$\lambda, \lambda_S
> 0$ and $\beta^2 <4 \lambda\lambda_S$ for $V$ to be positive-definitive.
In the unitary gauge, the charged and pseudoscalar neutral components of $H$
are absorbed by the $W^\pm$ and $Z$ gauge bosons, respectively, while the
pseudoscalar component of $S$ is absorbed by the new $Z'$. In the physical
basis, where
\begin{align}
H = \begin{pmatrix} 0\\ \dfrac{h+v}{\sqrt{2}} \end{pmatrix},\quad
S=\frac{s+v_S}{\sqrt{2}},
\end{align}
the potential has the form
\begin{align}
\begin{split}
	V &= \lambda v^2 h^2 + \lambda_S v_S^2 s^2 + \beta v v_S h s + \frac{1}{4}\lambda h^4
+\frac{1}{4}\lambda_S s^4\\
	&\quad + \lambda v h^3 +\lambda_S v_S s^3
+ \frac{1}{4}\beta h^2 s^2 + \frac{1}{2}\beta v h s^2 +  \frac{1}{2}\beta v_S h^2 s\,.
\end{split}
\end{align}
The mass matrix for the neutral scalars $h$ and $s$ is given by
\begin{align}
	\mathbf{M}^2 =
\begin{pmatrix}
2 \lambda v^2 & \beta v v_S\\ \beta v v_S & 2 \lambda_S v_S^2
\end{pmatrix},
	\label{eq:scalar_mass_matrix}
\end{align}
leading to the mass eigenstates $\phi_{1,2}$,
\begin{equation}
\begin{pmatrix}
\phi_1\\\phi_2
\end{pmatrix} =
\begin{pmatrix}
\cos \theta & - \sin \theta\\ \sin\theta & \cos
\theta
\end{pmatrix}
\begin{pmatrix}
h\\s
\end{pmatrix},
\end{equation}
with
\begin{equation}
	\tan 2\theta = \frac{\beta v v_S}{ \lambda_S v_S^2 - \lambda v^2}\,.
\end{equation}
The masses are
\begin{align}
m_{1,2}^2 = \lambda v^2 + \lambda_S v_S^2 \mp
\sqrt{(\lambda_S v_S^2 - \lambda v^2)^2 + \beta^2 v_S^2 v^2}\,.
\end{align}

In the limit $v_S \gg v$ and $\lambda_S v_S^2 \gg \lambda v^2$, one obtains
\begin{align}
m_1^2 \simeq 2
\left(\lambda - \frac{\beta^2}{4\lambda_S}\right) v^2, \quad m_2^2 \simeq
2 \lambda_S\, v_S^2\,,
\end{align}
and
\begin{equation}
\theta \simeq \frac{\beta v}{2 \lambda_S v_S}\,.
\end{equation}
The mass of the new $Z'$ gauge boson is given by  \mbox{$m_{Z'} = |x_s|\, g_X
v_S$}, where $g_X$ is the $U(1)_X$ gauge coupling.

An indirect constraint on $m_{Z'}$ comes from analyses of LEP2 precision
electroweak data~\cite{LEP:2003aa}:
\begin{align}
	\frac{m_{Z'}}{g_X} = |x_s|\,v_S \gtrsim 13.5~\text{TeV}.
\label{vSconstraint}
\end{align}
Thus, depending on the charge $x_s$, different lower bounds on the breaking
scale of the $U(1)_X$ gauge symmetry are obtained. For the anomaly-free
scalar charges given in Table~\ref{table2}, namely $|x_s|=2, 3, 6$, one
obtains the bounds $v_S \gtrsim 6.75$~TeV, 4.5~TeV, and 2.25~TeV,
respectively. To put limits on the $Z'$ mass, the gauge coupling strength
must be known. Assuming, for definiteness, $g_X \sim 0.1$, the bound in
Eq.~\eqref{vSconstraint} implies $m_{Z'} \gtrsim 1.4$~TeV. Such masses could
be probed through the search of dilepton $Z'$ resonances at the final stage
of the LHC, with a center-of-mass energy $\sqrt{s}=14$~TeV and integrated
luminosity $L \simeq 100$~fb$^{-1}$~\cite{Lee:2010hf,Godfrey:2013eta}. Recent
searches for narrow high-mass dilepton resonances at the LHC
ATLAS~\cite{ATLAS:2013jma} and CMS~\cite{CMS:2013} experiments have already
put stringent lower limits on extra neutral gauge bosons. In particular, from
the analysis of $pp$ collisions at $\sqrt{s}=8$~TeV, corresponding to an
integrated luminosity of about 20 fb$^{-1}$, these experiments have excluded
at 95\%~C.L. a sequential SM $Z'$ (i.e. a gauge boson with the same couplings
to fermions as the SM $Z$ boson) lighter than 3 TeV.

Electroweak precision data severely constrain any mixing with the ordinary
$Z$ boson~\cite{Langacker:2008yv}. The $Z-Z'$ mixing may appear either due to
the presence of Higgs bosons which transforms nontrivially under the SM gauge
group and the new $U(1)_X$ Abelian gauge symmetry or via kinetic mixing in
the Lagrangian~\cite{Holdom:1985ag}. The mass mixing is not induced in our
case because the SM Higgs doublet is neutral under $U(1)_X$, while kinetic
mixing may be avoided (up to one loop), if $U(1)_Y$ and $U(1)_X$ are
orthogonal~\cite{Loinaz:1999qh}. Although a detailed analysis of the $Z-Z'$
mixing is beyond the scope of our work, it is worth noting that, in general,
it imposes additional restrictions on these models. For simplicity, hereafter
we assume that mixing is negligible and restrict ourselves to the case with
no mixing.

\subsection{Nonstandard neutrino interactions}

Neutrino oscillation experiments are sensitive to new degrees of freedom that
mediate neutral-current interactions. The effects of new physics are usually
parametrized by the so-called nonstandard interactions (NSI), which affect
neutrino production and detection processes as well as neutrino propagation
in matter. In particular, NSI are generated in seesaw models once the heavy
(singlet or triplet) fermions are integrated out. In the canonical basis of
the kinetic terms, the induced $d=6$ operators lead to modified couplings of
the leptons to gauge bosons and, consequently, to deviations from unitarity
of the leptonic mixing matrix (see, e.g., Ref.~\cite{Branco:2011zb} and
references therein).

Nonstandard neutrino interactions are also induced in the presence of new
heavy gauge bosons with nonuniversal lepton couplings. These can be
parametrized in terms of the effective four-fermion operators
\begin{align}
\mathcal{L}_{\rm NSI}=-2\sqrt{2}G_F\, \varepsilon_{\alpha\beta}^{fP}\,
(\bar{\nu}_\alpha \gamma_\mu L \nu_\beta) (\bar{f}\gamma^\mu P f)\,,
\end{align}
where $G_F$ is the Fermi constant, $P=L$ or $R$ stands for the chiral
projection operator, $f$ denotes a SM fermion, ${\alpha,\beta}={e,\mu,\tau}$,
and $\varepsilon_{\alpha\beta}^{fP}$ are dimensionless couplings that encode
deviations from standard interactions. A simple estimate allows us to relate
the magnitude of the $\varepsilon$ parameters to the new physics scale.
Assuming that NSI are mediated by some intermediate particles with masses of
the order of $m_{\rm NSI}$, one expects $|\varepsilon| \sim m_{W}^2/m_{\rm
NSI}^2$, leading to $|\varepsilon| \sim 10^{-2}\, (10^{-4})$ for $m_{\rm NSI}
\sim 1\, (10)$~TeV.

In phenomenological studies of neutrino propagation in matter, it is
customary to define the effective NSI parameters
\begin{align}
\varepsilon_{\alpha\beta}=\sum_{f,P} \frac{n_f}{n_e}\,\varepsilon_{\alpha\beta}^{fP}\,,
\end{align}
where  $n_f$ is the number density of the fermion species $f=e, u, d$ in matter. For
neutral Earth-like matter the relation
\begin{align}
\varepsilon_{\alpha\beta} \simeq \sum_P\, \left(\varepsilon_{\alpha\beta}^{eP} +
3\varepsilon_{\alpha\beta}^{uP} + 3 \varepsilon_{\alpha\beta}^{dP}\right)
\end{align}
holds, while for neutral solar-like matter
\begin{align}
\varepsilon_{\alpha\beta} \simeq \sum_P\, \left(\varepsilon_{\alpha\beta}^{eP} +
2\varepsilon_{\alpha\beta}^{uP} +  \varepsilon_{\alpha\beta}^{dP}\right).
\end{align}
The above quantities can be approximately bounded by
\begin{align}\label{epsearth}
\varepsilon_{\alpha\beta}^\oplus \simeq \sqrt{\sum_P\,
\left[(\varepsilon_{\alpha\beta}^{eP})^2 + (3\varepsilon_{\alpha\beta}^{uP})^2 +
(3 \varepsilon_{\alpha\beta}^{dP})^2\right]}
\end{align}
and
\begin{align}\label{epssolar}
\varepsilon_{\alpha\beta}^\odot \simeq \sqrt{\sum_P\,
\left[(\varepsilon_{\alpha\beta}^{eP})^2 + (2\varepsilon_{\alpha\beta}^{uP})^2 +
(\varepsilon_{\alpha\beta}^{dP})^2\right]}\,,
\end{align}
for neutral Earth-like ($\oplus$) and solar-like ($\odot$)  matter,
respectively. The following model-independent bounds are then
obtained~\cite{Ohlsson:2012kf}:
\begin{align}\label{epsbounds}
\begin{split}
&|\varepsilon_{ee}^{\oplus}|<4.2\,,
\,|\varepsilon_{\mu\mu}^{\oplus}|<0.068\,,
\,|\varepsilon_{\tau\tau}^{\oplus}|<21.0;\\
&|\varepsilon_{ee}^{\odot}|<2.5\,,
\,|\varepsilon_{\mu\mu}^{\odot}|<0.046\,,
\,|\varepsilon_{\tau\tau}^{\odot}|<9.0.
\end{split}
\end{align}

\begin{table}[t]
\begin{tabular}{|c|c|c|c|c|c|}
\hline
$\;(x_e, x_\mu, x_\tau)\;$ & $\;|x_s|\;$ & $\;\;C_{\oplus}\;\;$ & $\;\;C_{\odot}\;\;$ &
\;$v^{\oplus}_S$ [GeV]\; & \;$v^{\odot}_S$ [GeV]\;\\
\hline
$(1,-1,-3)$ & 2 & $0.31$ & $0.22$ & $522$ & $539$ \\
$(3,-1,-5)$ & 2 & $0.59$ & $0.55$ & $723$ & $849$ \\
$(3,0,-6)$ & 3 & $0.26$ & $0.24$ & $106$ & $133$ \\
$(9,-3,-9)$ & 6 & $0.18$ & $0.18$ & $692$ & $837$ \\
$(1,-3,-1)$ & 2 & $0.31$ & $0.22$ & $905$ & $934$ \\
$(3,-5,-1)$ & 2 & $0.59$ & $0.55$ & $1617$ & $1898$ \\
$(3,-6,0)$ & 3 & $0.26$ & $0.24$ & $1181$ & $1386$ \\
$(9,-9,-3)$ & 6 & $0.18$ & $0.18$ & $1198$ & $1451$ \\
$(-1,1,-3)$ & 2 & $0.31$ & $0.22$ & $522$ & $539$ \\
$(-1,3,-5)$ & 2 & $0.31$ & $0.22$ & $905$ & $934$ \\
$(0,3,-6)$ & 3 & $0.11$ & $0.06$ & $545$ & $481$ \\
$(-3,9,-9)$ & 6 & $0.07$ & $0.06$ & $723$ & $849$ \\
$(-1,-3,1)$ & 2 & $0.31$ & $0.22$ & $905$ & $934$ \\
$(-1,-5,3)$ & 2 & $0.31$ & $0.22$ & $1168$ & $1205$ \\
$(0,-6,3)$ & 3 & $0.11$ & $0.06$ & $771$ & $680$ \\
$(-3,-9,9)$ & 6 & $0.07$ & $0.06$ & $723$ & $849$ \\
\hline
\end{tabular}
\caption{\label{tab:NSI}Lower bounds on the $U(1)_X$ breaking scale $v_S$,
assuming NSI of neutrinos in Earth-like matter ($v^{\oplus}_S$) and
solar-like matter ($v^{\odot}_S$). The bounds are given for each
anomaly-free solution presented in Table~\ref{table2}.}
\end{table}

For the anomaly-free models discussed in Sec.~\ref{sec:gaugesym} and
summarized in Table~\ref{table2}, the deviations from standard interactions
are given by
\begin{align}
\varepsilon_{\alpha\beta}^{fP}=\frac{v^2}{2 v_S^2}
\frac{x_f\, x_{\nu_\alpha}}{x_s^2} \delta_{\alpha\beta}.
\end{align}
Equations~\eqref{epsearth} and \eqref{epssolar} then imply
\begin{align}
\varepsilon_{\alpha\alpha}^m \simeq \frac{v^2}{v_S^2}\,
|x_{\nu_\alpha}|\, C_m,
\end{align}
with
\begin{align}
C_{\oplus}=\frac{1}{x_s^2} \sqrt{\frac{x_e^2}{2}+1}\,,\quad
C_{\odot}=\frac{1}{x_s^2} \sqrt{\frac{x_e^2}{2}+\frac{5}{18}}\,.
\end{align}
The limits given in Eq.~\eqref{epsbounds} can be translated into lower bounds
on the $U(1)_X$ gauge symmetry breaking scale $v_S$. For each anomaly-free
solution of Table~\ref{table2}, the corresponding bounds (obtained from NSI
of neutrinos in Earth-like and solar-like matter) are given in
Table~\ref{tab:NSI}. As can be seen from the table, these bounds are less
restrictive than those obtained from electroweak precision tests.

\subsection{Gauge sector and flavor model discrimination}

For the effects due to the new gauge symmetry to be observable, the seesaw
scale should be low enough. One expects a phenomenology similar to the case
with a minimal $B-L$ scalar sector~\cite{Basso:2010yz}. Nevertheless, by
studying the $Z'$ resonance and its decay products, one could in principle
distinguish the generalized $U(1)_X$ models from the minimal $B-L$ model.

Due to their low background and neat identification, leptonic final states
give the cleanest channels for the discovery of a new neutral gauge boson. In
the limit that the fermion masses are small compared with the $Z'$ mass, the
$Z'$ decay width into fermions is approximately given by
\begin{align}
 \Gamma (Z'\rightarrow f \overline{f}) \simeq \frac{g_X^2}{24\pi} m_{Z'}
 \left(x_{f L}^2+ x_{f R}^2\right),
\end{align}
where $x_{f L}$ and $x_{f R}$ are the $U(1)_X$ charges for the left and right
chiral fermions, respectively.

\begin{figure}[t]
\centering
\includegraphics[width=9cm]{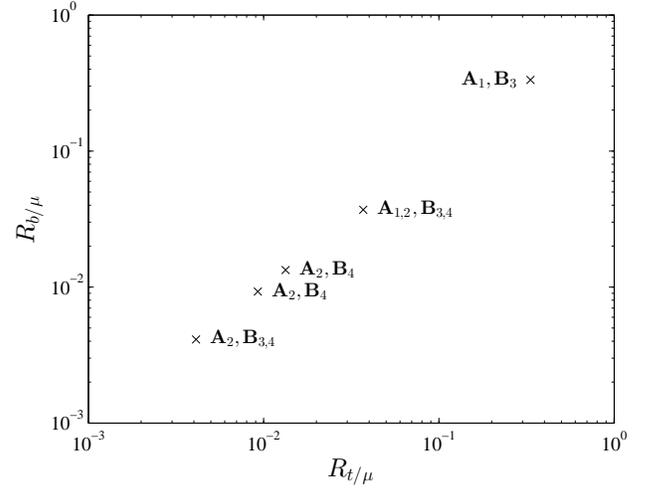}
\caption{\label{fig1} $R_{t/\mu}-R_{b/\mu}$ branching ratio plane for the
anomaly-free solutions of Table~\ref{table2}, leading to neutrino
mass matrix patterns of type $\mathbf{A}_{1,2}$ and $\mathbf{B}_{3,4}$.}
\end{figure}

It has been shown~\cite{Godfrey:2008vf,Diener:2010sy} that the decays of $Z'$
into third-generation quarks, $pp \rightarrow Z' \rightarrow b\,\overline{b}$
and $pp \rightarrow Z' \rightarrow t\,\overline{t}$ can be used to
discriminate between different models, having the advantage of reducing the
theoretical uncertainties. In particular, the branching ratios $R_{b/\mu}$
and $R_{t/\mu}$ of quarks to $\mu^+ \mu^-$ production,
\begin{align}\label{branchratios}
\begin{split}
    R_{b/\mu}&=\frac{\sigma(pp \rightarrow Z'
    \rightarrow b\,\overline{b})}{\sigma(pp
\rightarrow Z' \rightarrow \mu^+ \mu^-)} \simeq 3 K_{b}
\frac{x_q^2+x_{d}^2}{x_{\ell2}^2+x_{e2}^2}\,,\\
R_{t/\mu}&=\frac{\sigma(pp \rightarrow Z'
    \rightarrow t\,\overline{t})}{\sigma(pp
\rightarrow Z' \rightarrow \mu^+ \mu^-)} \simeq 3 K_{t}
\frac{x_q^2+x_{u}^2}{x_{\ell 2}^2+x_{e 2}^2}\,,
\end{split}
\end{align}
could serve as discriminators. The $K_{b,t} \sim \mathcal{O}(1)$ factors
incorporate the QCD and QED next-to-leading-order correction factors.
Substituting the quark and charged-lepton $U(1)_X$ charges, i.e.
$x_q=x_u=x_d=a/3$ and $x_{\ell 2}=x_{e 2}=-b_2$, we obtain
\begin{align}
R_{b/\mu} \simeq \frac{K_b}{3} \frac{a^2}{b_2^2}\,,
\quad R_{t/\mu}\simeq \frac{K_t}{3} \frac{a^2}{b_2^2}\,,
\end{align}
yielding $R_{b/\mu} \simeq R_{t/\mu}$. Figure~\ref{fig1} shows the
$R_{t/\mu}-R_{b/\mu}$ branching ratio plane for the anomaly-free solutions
given in Table~\ref{table2}, which lead to the viable neutrino mass matrix
patterns $\mathbf{A}_{1,2}$ and $\mathbf{B}_{3,4}$, with two independent
zeros. As can be seen from the figure, the solutions split into five
different points in the plane, which correspond to the allowed values of the
$b_2$ coefficient, $|b_2|=1, 3, 5, 6, 9$, assuming $a=1$. The allowed
$\mathbf{m}_\nu$ patterns are shown at each point.

\begin{figure}[t]
\centering
\includegraphics[width=9cm]{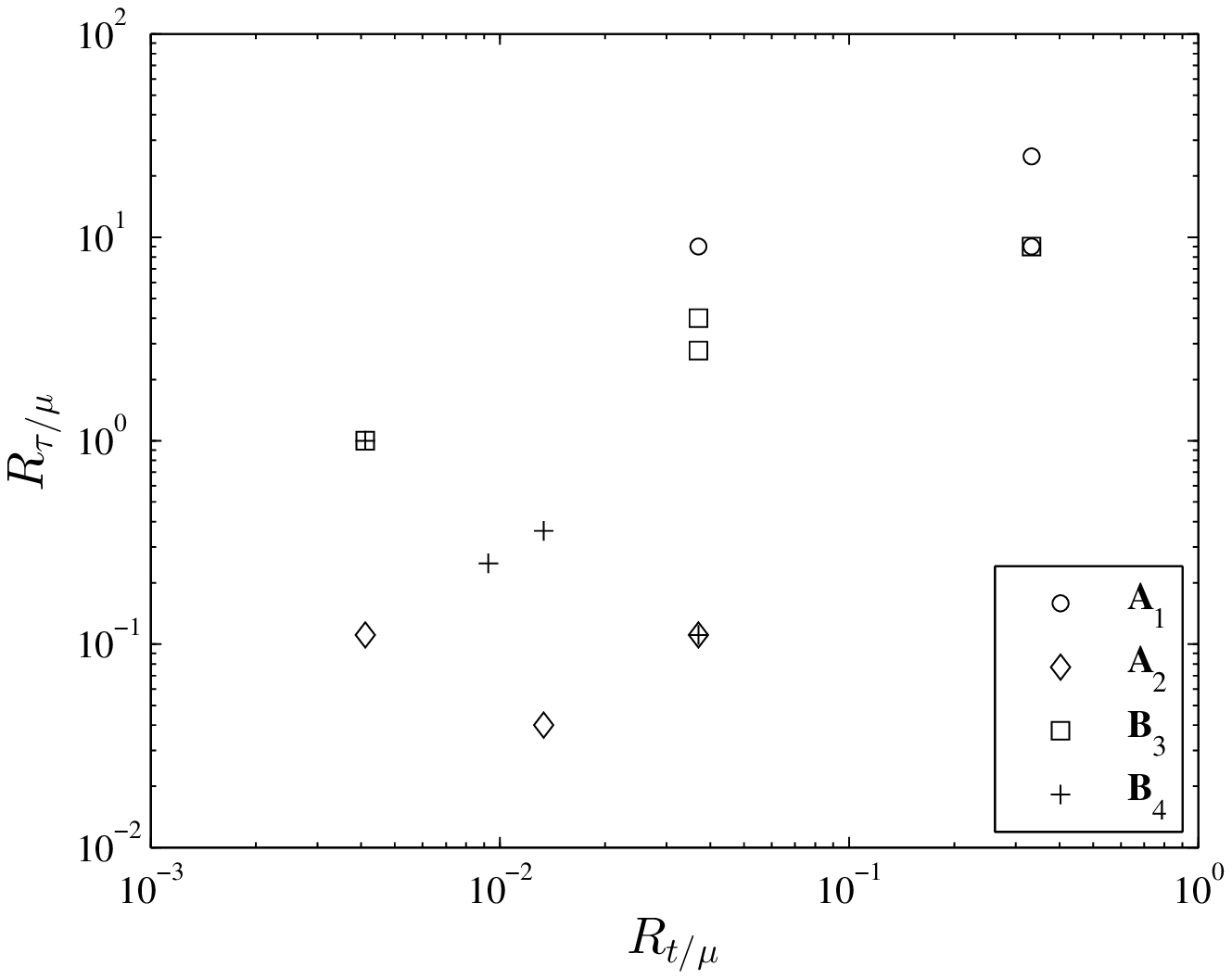}
\caption{\label{fig2} $R_{t/\mu}-R_{\tau/\mu}$ branching ratio plane for the
anomaly-free solutions of Table~\ref{table2}, leading to neutrino
mass matrix patterns of type $\mathbf{A}_{1,2}$ and $\mathbf{B}_{3,4}$.}
\end{figure}

The ratio $R_{\tau/\mu}$ of the branching fraction of $\tau^+\tau^-$ to
$\mu^+\mu^-$ has also proven to be useful for understanding models with
preferential couplings to $Z'$~\cite{Diener:2010sy}. It is approximately
given in our case by
\begin{align}\label{ratioRtaumu}
    R_{\tau/\mu}=\frac{\sigma(pp \rightarrow Z'
    \rightarrow \tau^+\tau^-)}{\sigma(pp
\rightarrow Z' \rightarrow \mu^+ \mu^-)} \simeq K_{\tau}
\frac{x_{\ell 3}^2+x_{e 3}^2}{x_{\ell 2}^2+x_{e 2}^2} \simeq K_\tau\,
\frac{b_3^2}{b_2^2}\,,
\end{align}
where in the last expression we have used the charge relation $x_{\ell
3}=x_{e 3}=-b_3$. Clearly, this ratio can be used to distinguish models with
generation universality ($R_{\tau/\mu} \simeq 1$) from models with
nonuniversal couplings, as those given in Table~\ref{table2}. The
$R_{t/\mu}-R_{\tau/\mu}$ branching ratio plane is depicted in
Fig.~\ref{fig2}. In this case, the neutrino mass matrix patterns exhibit a
clear discrimination in the plane, having overlap of two solutions in just
three points.

\section{Conclusions}
\label{sec:summary}

We have considered extensions of the SM based on Abelian gauge symmetries
that are linear combinations of the baryon number $B$ and the individual
lepton numbers $L_{e,\mu,\tau}$. In the presence of a type I and/or type III
seesaw mechanisms for neutrino masses, we then looked for possible charge
assignments under the new gauge symmetry that lead to cancellation of gauge
anomalies and, simultaneously, to a predictive flavor structure of the
effective Majorana neutrino mass matrix, consistent with present neutrino
oscillation data. Our analysis was performed in the physical basis where the
charged leptons are diagonal. This implies that the neutrino mass matrix
patterns with two independent zeros, obtained via the seesaw mechanism, are
directly linked to low-energy parameters. We recall that, besides three
charged lepton masses, there are nine low-energy leptonic parameters (three
neutrino masses, three mixing angles, and three CP violating phases).
Two-zero patterns in the neutrino mass matrix imply four constraints on these
parameters. Would we consider charge assignments that lead to nondiagonal
charged leptons, then the predictability of our approach would be lost, since
rotating the charged leptons to the diagonal basis would destroy, in most
cases, the zero textures in the neutrino mass matrix.

Working in the charged lepton flavor basis, we have found that only a limited
set of solutions are viable (cf. Table~\ref{table2}), leading to two-zero
textures of the neutrino mass matrix with a minimal extra fermion and scalar
content. All allowed patterns were obtained in the framework of the type I
seesaw mechanism with three right-handed neutrinos (or in a mixed type I/III
seesaw framework with three right-handed neutrinos and one fermion triplet),
extending the SM scalar sector with a complex scalar singlet field.

We have also studied how the nonuniversal $U(1)_X$ charge assignments to
neutrinos affect neutrino oscillation analysis through the matter effects. We
found that current model-independent limits on the effective NSI parameters
lead to lower bounds on the $U(1)_X$ symmetry breaking scale that are less
restrictive than those obtained from electroweak precision tests and, in
particular, from recent LHC experiments. Finally, we briefly addressed the
possibility of discriminating the different charge assignments (gauge
symmetries) and seesaw realizations at the LHC. We have shown that the
measurements of the ratios of third generation final states $(\tau,b,t)$ to
$\mu$ decays of the new gauge boson $Z'$ could be useful in distinguishing
between different gauge symmetry realizations. This provides a complementary
way of testing flavor symmetries and their implications for low-energy
neutrino physics.

\section*{Acknowledgements}
The work of D.E.C.~was supported by \emph{Associa\c c\~ao do Instituto
Superior T\'ecnico para a Investiga\c c\~ao e Desenvolvimento} (IST-ID) and
by \textit{Funda\c{c}\~{a}o para a Ci\^{e}ncia e a Tecnologia} (FCT) through
the projects PEst-OE/FIS/UI0777/2013, CERN/FP/123580/2011 and
PTDC/FIS-NUC/0548/2012. R.G.F. acknowledges support from FCT through the
projects PEst-OE/FIS/UI0777/2013 and CERN/FP/123580/2011.

\appendix

\section{Two-zero textures for the neutrino mass matrix and their seesaw realization}
\label{appexA}

\begin{table*}[t]
\begin{tabular}{|c|c|c|c|c|}
  \hline
    & & & & \\
  $\, \mathbf{M}_{R, \Sigma} \,$ & $\quad\mathbf{D}_2$\,, $\begin{pmatrix}
           0 & \ast & \ast \\
           \ast & \ast & 0 \\
           \ast & 0 & 0 \\
         \end{pmatrix}\quad$ &
         $\quad\mathbf{D}_1$\,, $\begin{pmatrix}
           0 & \ast & \ast \\
           \ast & 0 & 0 \\
           \ast & 0 & \ast \\
         \end{pmatrix}\quad$ &
         $\quad\mathbf{B}_4$\,, $\begin{pmatrix}
           \ast & \ast & 0 \\
           \ast & 0 & \ast \\
           0 & \ast & 0 \\
         \end{pmatrix}\quad$ &
         $\quad\mathbf{B}_3$\,, $\begin{pmatrix}
           \ast & 0 & \ast \\
           0 & 0 & \ast \\
           \ast & \ast & 0 \\
         \end{pmatrix}\quad$ \\
         & & & & \\  \hline
$\mathbf{m}_\nu$ & $\mathbf{A}_1$ & $\mathbf{A}_2$ & $\mathbf{B}_3$ & $\mathbf{B}_4$\\\hline
\end{tabular}
\caption{\label{table3}Viable type I (type III) seesaw realizations of
two-zero textures of the effective neutrino mass matrix $\mathbf{m}_\nu$ when
$n_R=3\ (n_\Sigma=3)$ and the Dirac-neutrino Yukawa mass matrix
$\mathbf{m}_D$ ($\mathbf{m}_T$) is diagonal, i.e.
$\mathbf{m}_{D,T}=\text{diag}\,(\ast,\ast,\ast)$. All cases belong to the
permutation set $\mathcal{P}_1$.}
\end{table*}

\begin{table*}[t]
\begin{tabular}{|c|cccc|c|}
  \hline
   $\,\, \mathbf{m}_{D,T}\,\, $ & \multicolumn{4}{|c|}{$\mathbf{M}_{R, \Sigma}$}
   & $\quad \mathbf{m}_\nu\quad $\\
  \hline
        & & & & &\\
  $\mathbf{A}_1$ & $\quad \mathbf{A}_1$\,, & $\begin{pmatrix}
           0 & 0 & \ast \\
           0 & \ast & 0 \\
           \ast & 0 & \ast \\
         \end{pmatrix}$\,, & $\begin{pmatrix}
           0 & 0 & \ast \\
           0 & \ast & \ast \\
           \ast &\ast & 0
         \end{pmatrix}$\,, &$\begin{pmatrix}
           0 & 0 & \ast \\
           0 & \ast & 0 \\
           \ast & 0 & 0
         \end{pmatrix}\quad$ &$\mathbf{A}_1$\\
              & & & & &\\ \hline
  & & & & &\\
  $\mathbf{A}_2$ & $\quad\mathbf{A}_2$\,, & $\begin{pmatrix}
           0 & \ast & 0 \\
           \ast & 0 & \ast \\
           0 & \ast & \ast \\
         \end{pmatrix}$\,, &$\begin{pmatrix}
           0 & \ast & 0 \\
           \ast & \ast & 0 \\
           0 & 0 & \ast \\
         \end{pmatrix}$\,, & $\begin{pmatrix}
           0 & \ast & 0 \\
           \ast & 0 & 0 \\
           0 & 0 & \ast \\
         \end{pmatrix}$ & $\mathbf{A}_2$\\
         & & & & &\\  \hline
                & & & & &\\
  $\mathbf{B}_3$ & $\quad \mathbf{B}_3$\,, & $\begin{pmatrix}
           \ast & 0 & 0 \\
           0 & 0 & \ast \\
           0 & \ast & \ast \\
         \end{pmatrix}$\,, & $\begin{pmatrix}
           \ast & 0 & \ast \\
           0 & 0 & \ast \\
           \ast &\ast & 0
         \end{pmatrix}$\,, &$\begin{pmatrix}
           \ast & 0 & 0 \\
           0 & 0 & \ast \\
           0 & \ast & 0
         \end{pmatrix}\quad$ &$\mathbf{B}_3$\\
              & & & & &\\ \hline
  & & & & &\\
  $\mathbf{B}_4$ & $\quad\mathbf{B}_4$\,, &$\begin{pmatrix}
           \ast & 0 & 0 \\
           0 & \ast & \ast \\
           0 & \ast & 0 \\
         \end{pmatrix}$\,, &$\begin{pmatrix}
           \ast & \ast & 0 \\
           \ast & 0 & \ast \\
           0 & \ast & 0 \\
         \end{pmatrix}$\,, &$\begin{pmatrix}
           \ast & 0 & 0 \\
           0 & 0 & \ast \\
           0 & \ast & 0 \\
         \end{pmatrix}$ &$\mathbf{B}_4$\\
         & & & & &\\  \hline
\end{tabular}
\caption{\label{table4}Viable type I (type III) seesaw realizations of two-zero
textures of $\mathbf{m}_\nu$ when $n_R=3\ (n_\Sigma=3)$ and assuming that
$\mathbf{m}_D$ ($\mathbf{m}_T$) belongs to
a permutation set $\mathcal{P}_i\,(i=1,2,3,4)$. Only
matrices $\mathbf{m}_D$ ($\mathbf{m}_T$) and $\mathbf{m}_\nu$
contained in $\mathcal{P}_1$ are allowed, sharing always the same pattern.}
\end{table*}

The neutrino mass matrix $\mathbf{m}_\nu$ is a symmetric matrix with six
independent complex entries. There are $6!/[n! (6 - n)!]$ different textures,
each containing $n$ independent texture zeros. One can show that any pattern
of $\mathbf{m}_\nu$ with more than two independent zeros ($n
> 2$) is not compatible with current neutrino oscillation data. For $n=2$,
there are fifteen two-zero textures of $\mathbf{m}_\nu$, which can be
classified into six categories ($\mathbf{A}, \mathbf{B}, \mathbf{C},
\mathbf{D}, \mathbf{E}, \mathbf{F}$):

\begin{align}
\mathbf{A}_1: ~
\begin{pmatrix} 0 & 0 & \ast \\ 0 & \ast & \ast \\
\ast & \ast &
\ast
\end{pmatrix}
\,, ~
\mathbf{A}_2: ~
\begin{pmatrix}
0 & \ast & 0 \\ \ast & \ast & \ast \\ 0 & \ast & \ast
\end{pmatrix}
\,;\nonumber
\end{align}
\begin{align}
\mathbf{B}_1: ~
\begin{pmatrix}
\ast & \ast & 0 \\ \ast & 0 & \ast \\ 0 & \ast & \ast
\end{pmatrix}\,, ~
\mathbf{B}_2: ~
\begin{pmatrix}
\ast & 0 & \ast \\ 0 & \ast & \ast \\ \ast & \ast & 0
\end{pmatrix}\,, \nonumber
\end{align}
\begin{align}
\mathbf{B}_3: ~
\begin{pmatrix}
\ast & 0 & \ast \\ 0 & 0 & \ast \\ \ast & \ast & \ast
\end{pmatrix}\,, ~
\mathbf{B}_4: ~
\begin{pmatrix}
\ast & \ast & 0 \\ \ast & \ast & \ast \\ 0 & \ast & 0
\end{pmatrix}\,;\nonumber
\end{align}
\begin{align}
\mathbf{C}: ~
\begin{pmatrix}
\ast & \ast & \ast \\ \ast & 0 & \ast \\ \ast & \ast & 0
\end{pmatrix}\,;\quad
\end{align}
\begin{align}
\mathbf{D}_1: ~
\begin{pmatrix}
 \ast & \ast & \ast \\ \ast & 0 & 0 \\ \ast & 0 & \ast
\end{pmatrix}\,, ~
\mathbf{D}_2: ~
\begin{pmatrix}
\ast & \ast & \ast \\ \ast & \ast & 0 \\ \ast & 0 & 0
\end{pmatrix}\,;\nonumber
\end{align}
\begin{align}
\mathbf{E}_1: ~
\begin{pmatrix}
0 & \ast & \ast \\ \ast & 0 & \ast \\ \ast & \ast & \ast
\end{pmatrix}\,, ~
\mathbf{E}_2: ~
\begin{pmatrix}
0 & \ast & \ast \\ \ast & \ast & \ast \\ \ast & \ast & 0
\end{pmatrix}\,, ~
\mathbf{E}_3:~
\begin{pmatrix}
0 & \ast & \ast \\ \ast & \ast & 0 \\ \ast & 0 & \ast
\end{pmatrix}\,;\nonumber
\end{align}
\begin{align}
\mathbf{F}_1: ~
\begin{pmatrix}
\ast & 0 & 0 \\ 0 & \ast & \ast \\ 0 & \ast & \ast
\end{pmatrix} \,, ~
\mathbf{F}_2: ~
\begin{pmatrix}
\ast & 0 & \ast \\ 0 & \ast & 0 \\ \ast & 0 & \ast
\end{pmatrix}\,, ~
\mathbf{F}_3: ~
\begin{pmatrix}
\ast & \ast & 0 \\ \ast & \ast & 0 \\ 0 & 0 & \ast
\end{pmatrix}\,;\nonumber
\end{align}
the symbol ``$\ast$" denotes a nonzero matrix element. In the flavor basis,
where the charged-lepton mass matrix $\mathbf{m}_l$ is diagonal and
$\mathbf{m}_l=\text{diag}\,(m_e,m_\mu,m_\tau)$, only seven patterns, to wit
$\mathbf{A}_{1,2}$, $\mathbf{B}_{1,2,3,4}$ and
$\mathbf{C}$~\cite{Frampton:2002yf}, are compatible with the present neutrino
oscillation data~\cite{Fritzsch:2011qv}.

\begin{table*}[t]
\begin{tabular}{|cc|c|c|}
  \hline
  \multicolumn{2}{|c|}{$\mathbf{m}_{D,T}$} & $\mathbf{M}_{R, \Sigma}$ &
  $\quad\mathbf{m}_\nu\quad$\\
  \hline
     & & &\multirow{4}{*}{$\mathbf{C}$ with}\\
  $\quad \begin{pmatrix}
           \ast & \ast \\
           0 & \ast \\
           \ast & 0 \\
         \end{pmatrix}\quad,$ & $\begin{pmatrix}
           \ast & \ast \\
           \ast & 0  \\
           0 & \ast
         \end{pmatrix}$ & $\begin{pmatrix}
           0 & \ast \\
           \ast & 0 \\
           \end{pmatrix}$ &\multirow{2}{*}{$\det\mathbf{C}=0$}\\
            & & &\\ \hline
       & & &\multirow{15}{*}{$\mathbf{C}$}\\
   $\quad \begin{pmatrix}
           \ast & \ast & \ast \\
           0 & 0 & \ast \\
           \ast & 0 & 0 \\
         \end{pmatrix}$\,, & $\begin{pmatrix}
           \ast & \ast & \ast \\
           \ast & 0 & 0  \\
           0 &0 & \ast
         \end{pmatrix}\quad$ & $\quad \begin{pmatrix}
           0 & 0 & \ast \\
           0 & \ast & 0 \\
           \ast & 0 & 0
         \end{pmatrix}\quad$ &\\
             & & &\\ \cline{1-3}
     & & &\\
  $\quad \begin{pmatrix}
           \ast & \ast & \ast \\
           0 & \ast & 0 \\
           \ast & 0 & 0 \\
         \end{pmatrix}$\,, & $\begin{pmatrix}
           \ast & \ast & \ast \\
           \ast & 0 & 0  \\
           0 & \ast & 0
         \end{pmatrix}\quad$ & $\quad \begin{pmatrix}
           0 & \ast & 0 \\
           \ast & 0 & 0 \\
           0 & 0 & \ast
         \end{pmatrix}\quad$ &\\
             & & &\\\cline{1-3}
      & & &\\
   $\quad \begin{pmatrix}
           \ast & \ast & \ast \\
           0 & 0 & \ast \\
           0 & \ast & 0 \\
         \end{pmatrix}$\,, & $\begin{pmatrix}
           \ast & \ast & \ast \\
           0 & \ast & 0  \\
           0 & 0 & \ast
         \end{pmatrix}\quad$ & $\quad \begin{pmatrix}
           \ast & 0 & 0 \\
           0 & 0 & \ast \\
           0 & \ast & 0
         \end{pmatrix}\quad$ &\\
             & & &\\ \hline

\end{tabular}
\caption{\label{table5}Viable type I (type III) seesaw realizations that lead
to the two-zero pattern $\mathbf{C}$ in $\mathbf{m}_\nu$. The cases with
$n_R=2\ (n_\Sigma=2)$ and $n_R=3\ (n_\Sigma=3)$ are displayed.}
\end{table*}

\begin{table*}[t]
\begin{tabular}{|c|c|cccc|c|}
  \hline
  $\mathbf{m}_{D}$ & $\mathbf{M}_{R}$ &
  \multicolumn{4}{|c|}{$\mathbf{m}_{T}, \mathbf{M}_{\Sigma}$} &
  $\quad\mathbf{m}_\nu\quad$\\
  \hline
  & & & & & &\\
    $\quad \begin{pmatrix}
           0 & 0 \\
           0 & \ast \\
           \ast & 0 \\
         \end{pmatrix}\quad$ & $\quad \begin{pmatrix}
           0 & \ast \\
           \ast & \ast \\
           \end{pmatrix}\quad$ & \multirow{7}{*}{$\quad \begin{pmatrix}
           0 & \ast \\
           \ast & 0 \\
           0 & 0 \\
         \end{pmatrix}$\,,} & \multirow{7}{*}{$\begin{pmatrix}
           \ast & \ast \\
           \ast & 0 \\
           \end{pmatrix}${\; \rm or}} & \multirow{7}{*}{$\;\begin{pmatrix}
           \ast & 0 \\
           0 & \ast \\
           0 & 0 \\
         \end{pmatrix}$\,,} & \multirow{7}{*}{$\begin{pmatrix}
           0 & \ast \\
           \ast & \ast \\
           \end{pmatrix}\quad$} &\multirow{7}{*}{$\mathbf{B}_1$}\\
  &  &  & & & &\\ \cline{1-2}
  &  &  & & & &\\
          $\quad \begin{pmatrix}
           0 & 0 \\
           \ast & 0 \\
           0 & \ast \\
         \end{pmatrix}\quad$ & $\quad \begin{pmatrix}
           \ast & \ast \\
           \ast & 0 \\
           \end{pmatrix}\quad$ & & & & &\\
           &  &  & & & &\\ \hline
  & & & & & &\\
    $\quad \begin{pmatrix}
           0 & 0 \\
           0 & \ast \\
           \ast & 0 \\
         \end{pmatrix}\quad$ & $\quad \begin{pmatrix}
           \ast & \ast \\
           \ast & 0 \\
           \end{pmatrix}\quad$ & \multirow{7}{*}{$\quad \begin{pmatrix}
           0 & \ast \\
           0 & 0 \\
           \ast & 0 \\
         \end{pmatrix}$\,,} & \multirow{7}{*}{$\begin{pmatrix}
           \ast & \ast \\
           \ast & 0 \\
           \end{pmatrix}${\; \rm or}} & \multirow{7}{*}{$\;\begin{pmatrix}
           \ast & 0 \\
           0 & 0 \\
           0 & \ast \\
         \end{pmatrix}$\,,} & \multirow{7}{*}{$\begin{pmatrix}
           0 & \ast \\
           \ast & \ast \\
           \end{pmatrix}\quad$} &\multirow{7}{*}{$\mathbf{B}_2$}\\
  &  &  & & & &\\ \cline{1-2}
  &  &  & & & &\\
          $\quad \begin{pmatrix}
           0 & 0 \\
           \ast & 0 \\
           0 & \ast \\
         \end{pmatrix}\quad$ & $\quad \begin{pmatrix}
           0 & \ast \\
           \ast & \ast \\
           \end{pmatrix}\quad$ & & & & &\\
           &  &  & & & &\\ \hline
\end{tabular}
\caption{\label{table6}Examples of type I/III mixed seesaw
realizations with two right-handed neutrinos and two fermion triplets
($n_R=n_\Sigma=2$) that lead to a neutrino mass matrix of type
$\mathbf{B}_{1,2}$. The solutions correspond to cases where the $3\times2$
Dirac-Yukawa mass matrices $\mathbf{m}_{D}$ and $\mathbf{m}_{T}$ contain the
maximum of allowed vanishing elements, i.e. four zeros.}
\end{table*}

We remark that, since the $U(1)_X$ gauge symmetry does not constrain the
values of the Yukawa couplings, any ordering of the charged leptons in the
flavor basis is allowed. Therefore, any permutation transformation acting on
the above patterns is permitted, provided that it leaves $\mathbf{m}_l$
diagonal. In particular, we find the following permutation sets:
\begin{align}\label{psets}
\begin{split}
&\mathcal{P}_1\equiv(\mathbf{A}_1, \mathbf{A}_2, \mathbf{B}_3,\mathbf{B}_4,
\mathbf{D}_1, \mathbf{D}_2),\\
&\mathcal{P}_2\equiv(\mathbf{B}_1, \mathbf{B}_2, \mathbf{E}_3), \\
&\mathcal{P}_3\equiv(\mathbf{C}, \mathbf{E}_1, \mathbf{E}_2),\\
&\mathcal{P}_4\equiv(\mathbf{F}_1, \mathbf{F}_2, \mathbf{F}_3).
\end{split}
\end{align}
Starting from any pattern belonging to a particular set, one can obtain any
other pattern in the same set by permutations.

We look for all possible type I and/or Type III seesaw realizations of
two-zero textures of the neutrino mass matrix $\mathbf{m}_\nu$ compatible
with the experimental data, i.e. that lead to a pattern $\mathbf{A}_{1}$,
$\mathbf{A}_{2}$, $\mathbf{B}_{1}$, $\mathbf{B}_{2}$, $\mathbf{B}_{3}$,
$\mathbf{B}_{4}$ or $\mathbf{C}$. We restrict our analysis to scenarios with
$n_R +n_\Sigma \leq 4$. Clearly, in our minimal theoretical framework, with
only one Higgs doublet and one singlet scalar, not all the solutions found
can be implemented in a theory free of anomalies under the gauge group
$U(1)_X$, with $X = a\, B - \sum_{i=1}^{n_G} b_i L_i$. In particular,
requiring charged leptons to be diagonal ($b_1 \neq b_2 \neq b_3$) imposes
strong constraints on the patterns of the Dirac-Yukawa mass matrices. Indeed,
only matrices $\mathbf{m}_D$ and $\mathbf{m}_T$ with one nonzero element per
column are permitted. In this appendix, nevertheless, we present the complete
set of solutions, regardless of their realization or not as anomaly-free
theories, since from the model building viewpoint these solutions may be of
interest in the context of extended theories or in the presence of discrete
symmetries.

In Table~\ref{table3}, we present all viable type I (type III) seesaw
realizations of two-zero textures of the effective neutrino mass matrix
$\mathbf{m}_\nu$ when $n_R=3\ (n_\Sigma=3)$ and the Dirac-neutrino Yukawa
mass matrix $\mathbf{m}_D$ ($\mathbf{m}_T$) is diagonal, i.e.
$\mathbf{m}_{D,T}=\text{diag}\,(\ast,\ast,\ast)$. All cases belong to the
permutation set $\mathcal{P}_1$. Table~\ref{table4} shows the viable type I
(type III) seesaw realizations of two-zero textures of $\mathbf{m}_\nu$ when
$n_R=3\ (n_\Sigma=3)$ and assuming that $\mathbf{m}_D$ ($\mathbf{m}_T$)
belongs to a permutation set $\mathcal{P}_i\,(i=1,2,3,4)$. Only matrices
$\mathbf{m}_D$ ($\mathbf{m}_T$) and $\mathbf{m}_\nu$ contained in
$\mathcal{P}_1$ are allowed, sharing always the same pattern, i.e. exhibiting
``parallel" structures.

One may wonder whether neutrino mass matrices of type $\mathbf{C}$, belonging
to the permutation set $\mathcal{P}_3$, can be obtained. As it turns out,
this requires matrices $\mathbf{m}_{D,T}$ (and $\mathbf{M}_{R,\Sigma}$) with
two and four zeros, for $n_{R,\Sigma}=2$ and $n_{R,\Sigma}=3$, respectively.
In Table~\ref{table5} we present all viable type I (type III) seesaw
realizations that lead to the two-zero pattern $\mathbf{C}$ in the effective
neutrino mass matrix $\mathbf{m}_\nu$. The cases with $n_R=2\ (n_\Sigma=2)$
and $n_R=3\ (n_\Sigma=3)$ are considered. As can be seen from the table, with
only two right-handed singlet (fermion triplet) neutrinos, there are only two
possible constructions, both leading to a massless neutrino ($\det
\mathbf{C}=0$). In fact, these are the only solutions that yield a pattern
consistent with neutrino oscillation data; no textures of type $\mathbf{A}_i$
or $\mathbf{B}_i$ are found. For $n_R=3\ (n_\Sigma=3)$, besides the
$\mathbf{C}$-pattern, there exist several combinations of matrices
$\mathbf{m}_{D,T}$ and $\mathbf{M}_{R,\Sigma}$ (not displayed in the table),
that lead to the viable patterns $\mathbf{A}_{1,2}$ and $\mathbf{B}_{3,4}$.
However, no texture of type $\mathbf{B}_{1,2}$, belonging to the permutation
set $\mathcal{P}_2$, can be obtained. The latter can be realized in the
context of mixed seesaw schemes. In Table~\ref{table6}, several patterns
leading to neutrino mass matrices of type $\mathbf{B}_{1,2}$ through a mixed
seesaw with two right-handed neutrinos and two fermion triplets
($n_R=n_\Sigma=2$) are shown. The solutions correspond to cases where the
Dirac-Yukawa mass matrices $\mathbf{m}_{D,T}$ contain the maximum of allowed
vanishing matrix elements, i.e. four zeros. We remark that in the mixed cases
with $n_{R}=2, n_{\Sigma}=1$ and $n_{R}=1, n_{\Sigma}=2$ there are viable
patterns as well, but since they only generate neutrino mass matrices of type
$\mathbf{A}_{1,2}$, $\mathbf{B}_{3,4}$ and $\mathbf{C}$, we do not present
them in Table~\ref{table6}.


%

\end{document}